\documentclass[pra,aps,eqsecnum,twocolumn,showpacs]{revtex4}

\usepackage{graphicx}
\usepackage{amssymb,amsmath}

\begin{document}

\title{Quantum beat spectroscopy: stimulated emission probe of hyperfine quantum beats in the atomic Cs $8p\,^2P_{3/2}$ level}

\author{S.B. Bayram${}^{1}$, P. Arndt${}^{1}$, O.I. Popov${}^{1}$, C. Guney${}^{2}$, W.P. Boyle${}^{1}$, M.D. Havey${}^{3}$, and J. McFarland${}^{1}$}%
\affiliation{\small $^{1}$Department of Physics, Miami University, Oxford, Ohio 45056}%
\affiliation{\small $^{2}$Physics Department, Faculty of Science, Istanbul University,
Istanbul, Turkey}
\affiliation{\small $^{3}$Department of Physics, Old Dominion University,
Norfolk, VA 23529}

\email{bayramsb@MiamiOH.edu}

\date{\today}

\sloppy

\begin{abstract}
Measurements of hyperfine polarization quantum beats are used to determine the magnetic dipole (A) and electric quadrupole (B) coupling constants in the excited atomic Cs 8p $^{2}P_{3/2}$ level.  The experimental approach is a novel combination of pulsed optical pumping and time-delayed stimulated emission probing of the excited level.  From the measured evolution of the atomic linear polarization degree as a function of probe delay time, we determine the hyperfine coupling constants A = 7.42(6) MHz and B = 0.14(29) MHz.

\end{abstract}


\maketitle

\section{Introduction}
Determinations of atomic and molecular structural properties have a long tradition in physical sciences.  Some quantities, such as atomic energy level positions and ground state hyperfine coupling constants can, and have been, experimentally determined to extraordinary precision.   On the other hand, excited state properties such as, for instance, excited-state hyperfine coupling constants, multipole matrix elements or other parameters that depend on such matrix elements (such as atomic polarizabilities), can normally be measured to much lower precision.  For example, atomic dipole matrix elements have been measured in the best cases to only several parts in $10^{-4}$ \cite{mcal1,mcal2,oates,volz1,volz2,tanner,rafac1,young,rafac2,rafac3,vasilyev,ekstrom,jones,wang,beger,meyer,havey,bayram,markhotok,herold,bouloufa}.

Over the past decade, there have been two important motivating factors for more precise determinations of atomic dipole matrix elements.  One of these is the need to extract weak interaction coupling constants from atomic physics parity violation measurements in atomic cesium and francium \cite{derevianko1,wood,bouchiat,khriplovich,ginges,Johnson1}.  Another is the need in precision measurements for so-called magic-wavelength optical dipole traps \cite{katori,derevianko,safronova}. In such traps, the trap-induced light shift is the same for two states of the system under study, permitting long and nearly perturbation-free measurement times, among other things.  In each of these cases, calculating the necessary atomic quantities for a given atomic system requires a large number of very-well-known dipole matrix elements.  For this reason, precise measurements of accessible atomic parameters is essential to provide fiducial or benchmark values to assess the reliability of theoretical approaches used to calculate them. Benchmark values of matrix elements provide one measure of such reliability.   Similarly, measurements of atomic hyperfine coupling constants, which provide information on  nuclear charge or current distributions, also give a measure of electronic wave functions at small distances from the nucleus, where relativistic effects on the wave functions are quite influential.

In this paper we describe and apply an extension of a pump-probe hyperfine quantum beats approach developed earlier \cite{Bayram2008,NaHFQB,KHFQB1,KHFQB2,CsHFQB}.  In previous measurements, a short-pulse linearly polarized pump beam generated an initial value of electronic alignment components $<A_q>$ in an excited atomic level. Here the subscript $q$ is a component index.  The alignment components evolve in time according to $<A_q>$$g^{(2)}(t)$, where $g^{(2)}(t)$ is a time-dependent coefficient that contains the hyperfine coupling constants. Measurement of time dependence of $<A_q>$$g^{(2)}(t)$ allows extraction of excited state hyperfine coupling constants at a relative precision $~$ $10^{-4}$, even in cases where the hyperfine splittings are masked by the natural width of the excited level.  Previously, $<A_q>$$g^{(2)}(t)$ was experimentally determined by time-delayed and polarization-dependent probing with a short-pulse probe laser tuned to a more energetic excited state level.  Although this approach works well in many cases, it becomes increasingly difficult to generate an optical probe when the probed level has to be close to the ionization level of the atom under study.  One of several possible approaches to overcoming this limitation, and one that we introduce here, is to probe the time-dependent alignment by stimulated emission probing to a lower energy level.  With this approach, it is possible to access energetically high lying atomic or molecular level, and yet probe via optical optical double resonance to a level that provides a convenient fluorescence channel for detection.

The remainder of this report is organized as follows.  We first describe the experimental approach, including the general scheme, and experimental details as necessary.  This is followed by the details of the analysis required to extract the hyperfine coupling constants.  The measurement and analysis results are presented in the next section, along with comparisons with earlier measurements of the hyperfine coupling constants for the atomic Cs $8p\,^2P_{3/2}$ level.

\section{General Experimental Approach}

\subsection{Experimental Arrangement}
In this section we give a brief overview of the experimental scheme and arrangement that was used to perform the experiments.  The basic scheme is illustrated by the partial energy level diagram for $^{133}$Cs in Figure 1.  From the figure, we refer to resonant laser excitation of the $8p\,^2P_{3/2}$ level as the pump transition, and subsequent stimulated emission on the $8p\,^2P_{3/2}$ $\rightarrow$ $5d\,^2D_{5/2}$ as the probe transition. Cascade fluorescence from the $5d\,^2D_{5/2}$ level populates the $6p\,^2P_{3/2}$  level nearly exclusively, which subsequently decays by spontaneous emission to the $6s\,^2S_{1/2}$  level.  This decay channel then serves as a convenient monitor of the initial excitation process.

\begin{figure}[th]
\begin{center}
{$\scalebox{1.20}{\includegraphics*{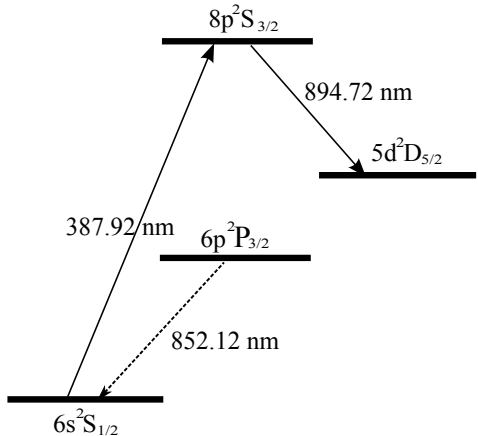}}$ }
\caption{Partial energy level diagram for atomic Cs, showing the excitation (solid line) and detection transitions (dotted line) used in the experiment. }
\end{center}
\par
\label{fig1}
\end{figure}

Tunable dye lasers to excite the pump and probe transitions are pumped by a pulsed neodymium-doped yttrium aluminum garnet (Nd:YAG) laser, operating simultaneously at 532 nm and at 355 nm.  The pulse repetition rate is 20 Hz.   This laser is used to drive two home-built tunable dye lasers, the so-called pump and probe lasers. The pump laser is used for excitation of the $8p\,^2P_{3/2}$ level at 387.92 nm and the probe laser is used to stimulate emission from the $8p\,^2P_{3/2}$ level to the $5d\,^2D_{5/2}$ level at 894.72 nm.  Each dye laser is highly linearly polarized through use of Glan-Thompson calcite
prism polarizers having extinction ratios of better than 10$^{-5}$.  The dye laser cavities are of the grazing incidence Littman-Metcalf design.  A temperature-controlled liquid crystal variable retarder (LCR) is used to electronically vary the
linear polarization direction of the probe laser to be parallel or perpendicular to that of the pump laser. The average power of the lasers is $\sim$ 1 mW and lasers are collimated to a beam diameter of about 0.3 cm. Polarization switching of the LCR was achieved by applying the necessary voltage to the retarder via a
computer-controlled liquid crystal digital interface.  The beams of the pump and probe pulsed dye lasers propagate collinearly, but in opposite directions, into the interaction region of the cesium sample cell.  A resistively heated
nonmagnetic cylindrical aluminum oven was used to generate the desired vapor pressure of atomic Cs in the cell. The oven, which houses the Pyrex cell containing cesium vapor, was wrapped with an aluminum oxide blanket to maintain the temperature at $70^{o}$C with the uncertainty under $\pm 0.01^{o}$C via a temperature controller.

\begin{figure}
\begin{center}
{$\scalebox{1.0}{\includegraphics*{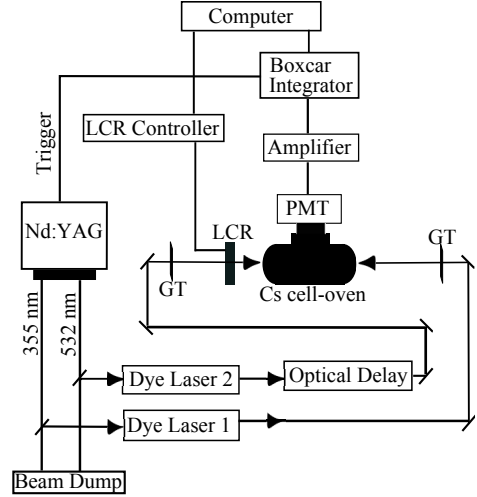}}$ }
\caption{A schematic view of the experimental apparatus.  In the figure GT stands for a Glan-Thompson polarizer, PMT for photomultiplier tube, and LCR refers to liquid crystal retarder.}
\end{center}
\par
\label{fig1}
\end{figure}
The intensities of the cascade fluorescence from the $6p\,^2P_{3/2}$ level to the ground $6s\,^2S_{1/2}$ level were recorded at 852.12 nm by using an infrared sensitive cooled photomultiplier tube (PMT) which was located at right angles to the propagation directions of the lasers. A combination of interference and color glass filters was used in front of the PMT in order to remove background light. All the cables used in the experiment were electrically shielded and the optical table was grounded in order to eliminate electronic pick-up and noise on the observed signal.

The recorded signal collected for each state of laser polarization consists of 100x10$^6$ data points
accumulated during 4 seconds. The boxcar integrator/averager was set to average every 100 data samples, where
each sample of the data set is continuous over a 50 ms time period. Since the lifetimes of the
$5d\,^2D_{5/2}$ and $6p\,^2P_{3/2}$ levels are shorter than the lifetime of the $8p\,^2P_{3/2}$ level (305 ns~\cite{Rad85})
we distinguished the signal from the background which consists of the spontaneous emission decay of the
atoms from the $8p\,^2P_{3/2}$ level to the lower levels.

The output of the PMT was amplified using a two-stage amplifier and
processed in the boxcar integrator/averager with a 50-ns gate-width, opened after a 10 ns
delay following the laser pulses. The boxcar integrator operated in a 30 sample averaging mode,
where the average single-shot level within the detection gate is digitized. Our typical
signal size is about 10$^{3}$ photons for each laser pulse.
The digitized signals were stored on a computer using a LabVIEW program while
monitoring the size of the signal within the gate-width in real time
using a digital oscilloscope operating at 500 MHz with 2 GSa/s. Comparison of the signals,
detected when the probe polarization angle is $\chi$ = 0 versus $\chi$ = $\pi$/2 allows definition of a linear polarization degree.  In particular, from the measurements, a linear polarization degree is formed from the measured intensities I($\chi$=0) and I($\chi$ = $\pi$/2) according to

\begin{equation}
P_L=\frac{I(\chi = 0) - I(\chi = \pi /2)}{I(\chi = 0) + I(\chi = \pi /2)}
 \label{Eq1n}
 \end{equation}

$P_L$, whose value can depend on the hyperfine energy separations in the probed $8p\,^2P_{3/2}$ level, is the main quantity to be further analyzed in the experiment.  As we see in the following section $P_L$ depends experimentally on the time delay between the pump and probe laser beams.  This time delay is generated by construction of an optical delay line, from which the geometrical properties yield a variable time delay.  Although not very convenient for long delays, an optical delay line has the advantage of generating a time base that is both of the necessary precision and largely free of important systematic errors. Linear polarization measurements depend mainly on the absolute intensity ratio of the signals for two different polarization directions of the lasers, and are quite insensitive to other experimental factors. Therefore, any variations of the laser intensities with experimental factors such as absorbing medium density, fluorescence background, and sensitivity of the gated boxcar averager generally have negligible effect on the intensity ratio. Laser power and temperature tests show that the amplified spontaneous emission, stimulated Raman scattering, radiation trapping or other phenomena do not affect the measured linear polarization.

\subsection{Experimental Analysis}
In the previous section we defined a linear polarization degree in terms of measured quantities. Here we sketch the necessary theoretical results with which the the experimental measurements are to be further analyzed to obtain the hyperfine coupling constants associated with the atomic Cs $8p\,^2P_{3/2}$ level.

In the present experiment, excitation of the atomic Cs $8p\,^2P_{3/2}$ level is made with linearly polarized light tuned to the $6s\,^2S_{1/2}$$\rightarrow$$8p\,^2P_{3/2}$ level. Further, the bandwidth of the optical excitation is much larger than the hyperfine splitting in the final level.  This means that the excited level may be characterized by an overall population and the axially symmetric electronic alignment tensor component $<A_o>$.  Thus, measured intensities in Eq.~2.1 can be defined in terms of alignment and general analysis of this situation~\cite{zaregreene} yields the following expression for the linear polarization degree associated with the stimulated emission probing described in the previous section.

\begin{equation}
P_L=\frac{3h^{(2)}\left(J_i,J_f\right)
\langle{A_o}\rangle}{4+h^{(2)}\left(J_i,J_f\right)
\langle{A_o}\rangle}
 \label{Eq1}
 \end{equation}

We described in detail the definition of the intensity of emitted radiation in terms of average electronic alignment in our earlier two-photon polarization spectroscopy experiments~\cite{Bayram2012,Bayram2009}. In the above expression, $h^{(2)}$ is a coefficient which is unique to each optical transition from an electronic level of atomic angular momentum $J_i$ to $J_f$.  In the present case, $J_i$ = 3/2 and $J_f$ = 5/2, these associated with the stimulated transition $8p\,^2P_{3/2}$ $\rightarrow$ $5d\,^2D_{5/2}$.  Here, $h^{(2)}$ = -1/4. Further, we create by optical excitation on the $6s\,^2S_{1/2}$ $\rightarrow$ $8p\,^2P_{3/2}$ transition an initial value of electronic alignment $\langle{A_o}\rangle$ = -4/5.  In the absence of further disturbance this quantity evolves in time according to $<A_o(t)>$ = $g^{(2)}(t)<A_{o}(0)>$, where the quantity $g^{(2)}(t)$ is given by the following expression:

\begin{eqnarray}\label{Eq2}
    g^{\left(2\right)}\left(t\right) =
    \sum_{F,F'}{\frac{\left(2F+1\right)\left(2F'+1\right)}{\left(2I+1\right)}\left\{
    \begin{array}{ccc}
    F & F' & 2 \\
    J_i & J_i & I \\
    \end{array}
    \right\}^2
    cos\left(w_{F,F'}t\right)},
\end{eqnarray}

\noindent where $t$ is the time delay between creation of the alignment, the
symbol $\{\cdots\}$ is a 6-j coefficient, $w_{F,F'}$ is the hyperfine frequency of the splitting between the $F$ and $F'$ hyperfine energy sublevels. The nuclear spin of $^{133}Cs$ is $I$ = 7/2. For the present case, the theoretical expression for the linear polarization degree is given by

\begin{equation}
P_L\left(t\right)=\frac{3g^{\left(2\right)}\left(t\right)}{20+
g^{\left(2\right)}\left(t\right)}.
 \label{Eq3}
 \end{equation}

Here the time dependent depolarization coefficient $g^{(2)}(t)$ is given by

\begin{eqnarray}\label{Eq4}
g^{(2)}(t) = 0.2187 &+& 0.09375\ cos\left[2\pi\nu_{23}t\right] \\ \nonumber &+& 0.2009\
cos\left[2\pi\nu_{24}t\right]
\\ \nonumber
&+& 0.0375\ cos\left[2\pi\nu_{34}t\right] \\
\nonumber &+& 0.16042\ cos\left[2\pi\nu_{35}t\right] \\ \nonumber &+& 0.28875\ cos\left[2\pi\nu_{45}t\right].
\end{eqnarray}

In terms of the magnetic dipole and electric quadrupole coupling constants A and B, the hyperfine frequencies to be inserted in this equation are given by

\begin{subequations}\label{Eq5}
\begin{eqnarray*}
           \nu_{23} &=& 3 A\ -\ \frac{5}{7} B \\
           \nu_{24} &=& 7 A\ -\ B \\
           \nu_{34} &=& 4 A\ -\ \frac{2}{7} B \\
           \nu_{35} &=& 9 A\ +\ \frac{3}{7} B \\
           \nu_{45} &=& 5 A\ +\ \frac{5}{7} B.
         \end{eqnarray*}
\end{subequations}

We finally point out that at $t$ = 0, the depolarization coefficient $g^{(2)}(0)$ = 1.   Then the ideal linear polarization degree for very short times following excitation is given by $P_L$ = 1/7 (14.29\%).

\section{Results and discussion}
As described in the previous section, the basic quantities obtained in the experiment are light intensities for the two cases when the exciting and probing linear polarization degrees are collinear or orthogonal.  These quantities, along with their associated uncertainties, are measured as function of the time delay between the pump and probe lasers.   The measured values are summarized in Table 1.

\begin{table}[h!]
  \centering
   \label{Table1}
   \caption{Tabulation of measured time delay between the pump and the probe beams, the measured linear polarization degrees, and the associated estimated uncertainties. The estimated uncertainty in the time delay is about 0.16 ns for each time delay.}
  \begin{tabular}{c c c c}
    \hline
    \hline
    \# & $t$ (ns)   & $P_L$ (\%) & $\Delta P_L$ (\%)
    \\ \hline
1   &   0.9      &   13.0  &   1.0    \\
2   &   1.8      &   12.4  &   1.4    \\
3   &   2.0      &   11.5  &   3.0    \\
4   &   2.8      &   11.9  &   1.4     \\
5   &   4.1      &   3.3   &   2.4    \\
6   &   7.1      &   0.0   &   0.6    \\
7   &   8.1      &   -2.4  &   1.4    \\
8   &   10.2     &   -3.2  &   1.9    \\
9   &   12.2     &   -2.3  &   0.3    \\
10  &   14.3     &   2.0   &   1.0    \\
11  &   17.3     &   3.3   &   2.6    \\
12  &   20.4     &   4.5   &   1.6    \\
13  &   22.4     &   2.2   &   2.9    \\
14  &   25.4     &   4.1   &   0.7    \\
15  &   27.5     &   4.9   &   1.8    \\
16  &   29.5     &   4.3   &   0.7    \\
17  &   30.0     &   5.2   &   2.5     \\
18  &   38.0     &   1.4   &   1.2    \\
19  &   43.0     &   2.8   &   1.0    \\
20  &   49.5     &   3.2   &   1.1     \\
21  &   59.0     &   8.8   &   1.9    \\
22  &   62.1     &   2.8   &   0.5    \\
23  &   65.1     &   -4.5  &   1.7    \\
24  &   68.2     &   -7.4  &   0.6    \\
25  &   71.2     &   -0.1  &   0.6    \\
26  &   74.3     &   5.6   &   1.0    \\
27  &   77.4     &   9.7   &   2.0    \\
28  &   80.4     &   7.5   &   0.4    \\
29  &   83.5     &   3.2   &   0.5    \\
30  &   86.5     &   3.1   &   1.7    \\
31  &   101.8    &   4.2   &   0.9    \\
32  &   104.8    &   6.5   &   0.9    \\
33  &   107.9    &   5.2   &   2.4    \\
34  &   109.9    &   3.2   &   0.2    \\
35  &   113.0    &   2.8   &   1.4    \\
36  &   114.0    &   2.4   &   4.0    \\
37  &   117.0    &   2.8   &   0.8    \\

    \hline
    \hline

  \end{tabular}
 \end{table}

The data in Table I are fitted using approaches developed earlier \cite{NaHFQB,KHFQB1,KHFQB2,CsHFQB} to model polarization-dependent hyperfine quantum beats in pump-probe transition.   The basic theoretical results we employ here are Eq. 2.3 - 2.4 as given in the previous section.  The fitting is done so as to minimize the reduced chi-squared of the fit, defined as

\begin{equation}\label{eq:chi2}
    \chi^2=\sum_i{\frac{\left[P^i_L\left(fit\right)-P^i_L\left(measured\right)\right]^2}{\eta
    \sigma^2_i}}.
\end{equation}

In this equation $P^i_L(fit)$ is the fitted linear polarization degree and $P^i_L(measured)$ is the measured value for the \textit{i}-th data point. The index \textit{i} refers to the first column in Table I. $\sigma^2_i$ is the squared estimated uncertainty of the \textit{i}-th data point, while $\eta$ is the number of degrees of freedom in the fit.  The fitting parameters are the magnetic dipole coupling constant A, the electric quadrupole coupling constant B, the temporal width W of the pump and probe laser beams, and an overall time offset $\delta$t between the pump and probe beams.  The parameter W is required because of the finite temporal width of the pump and probe beams in comparison with the inverse oscillation frequencies of the hyperfine quantum beats.   This is modeled by considering each laser pulse to be temporally rectangular.  This introduces a multiplicative quantity for each oscillating term in the alignment $<A_o>$ given by $2[1-cos(\omega_{FF'}W)]/[\omega_{FF'}W]^2$.  Here $\omega_{FF'}$ is the angular frequency separation between the hyperfine levels $F$ and $F^{\prime}$.  Previous studies have shown that this relatively small effect is not sensitive to the model of the laser temporal pulse shape \cite{NaHFQB,Bayram94}.  The overall temporal offset $\delta$t is to account for the fact that the pulsed lasers have relative starting times that are not directly measured in the experiment.  The time delays given in Table I represent the time delay given by the geometrical difference in path lengths between the pump and the probe beams.  The values of the parameters that yielded a minimum reduced chi squared are summarized in Table II.

\begin{table}
\caption{Parameters obtained by fitting the experimental data to the theoretical expressions. The magnetic dipole coupling constant A, the electric quadrupole coupling constant B, the temporal width W of the pump and probe laser beams, and an overall time offset $\delta$t between the pump and probe beams are the fitting parameters. Note that the A and B coefficients for this work have assigned 2$\sigma$ error bars.}
\begin{tabular}{c c c c}
  \hline
  \hline
  A(MHz) & B(MHz) & $\delta$t(ns) & W(ns) \\
  \hline
  7.42(6) & +0.14(29) & 0.02(52) & 2.4(3.3) \\

  \hline
  \hline

\end{tabular}
\end{table}

The data from Table I is plotted as a function of delay time $t$ in Figure 2, along with the result of the fit.  It can be seen in the figure that the qualitative agreement between the measured values and those calculated using the parameters in Table II is quite satisfactory.

\begin{figure}[th]
\begin{center}
{$\scalebox{0.27}{\includegraphics*{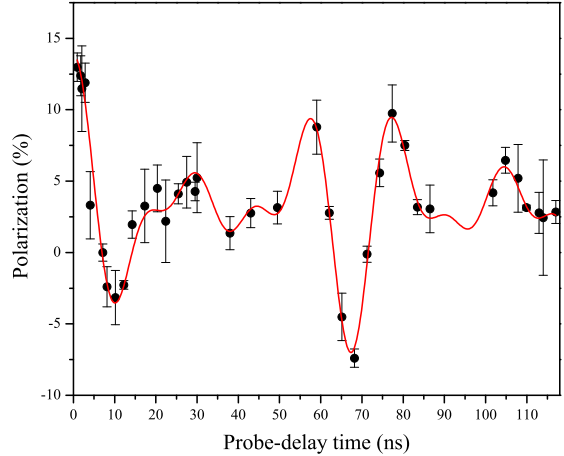}}$ }
\caption{Quantum beats are experimentally observed from the pump-stimulated emission probe spectroscopy. The probe-delay time dependence of the linear polarization degree showing the hyperfine quantum beats in the excited level is indicated by the data points.  The result of the best fit to the data is the solid line.}
\end{center}
\label{fig2}
\end{figure}

To further illustrate the quality of the fit, we present in Figure 3 the residuals of the fit as a function of delay time $\delta$t.  There is seen very good overall agreement with an expected Gaussian statistical distribution centered near an average value of zero; the spread in the residuals is also consistent with this distribution, with approximately $86 \%$ of the data points within the $\pm$ 1-$\sigma$ band.

\begin{figure}[th]
\begin{center}
{$\scalebox{0.27}{\includegraphics*{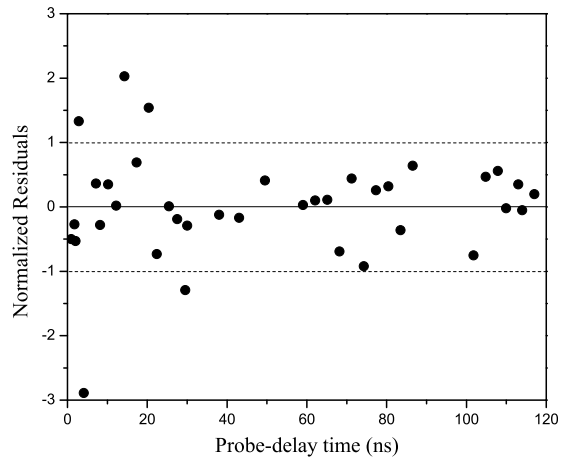}}$ }
\caption{Distribution of the residuals of the fit of the linear polarization as a function of probe-delay time. Each residual is normalized to the experimental estimated uncertainty in the individual data point. The solid horizontal line represents the average value of the residuals, while the dashed horizontal lines indicate the normalized one-$sigma$ levels.}
\end{center}
\label{fig3}
\end{figure}

In Table III we present a comparison of our experimentally determined hyperfine coupling constants with those obtained previously by other methods.  As can be seen, there are relatively few theoretical or experimental determinations of the hyperfine structure of the atomic Cs $8p\,^2P_{3/2}$ level. This is somewhat surprising, considering how extensively the various structural properties of the Cs atom have been researched in the context of atomic parity violation studies \cite{derevianko1,wood,bouchiat,khriplovich,ginges,Johnson1}.  The distribution of the coupling constants A and B for previous measurements is also quite wide, especially in comparison with measurements on many other alkali atom levels \cite{Arimondo}.
Although we have no specific explanation for the relatively wide range of the hyperfine coupling constants given in Table III, we believe our results to be solid, and to be free of many of the systematic errors commonly found in double resonance measurements.   In particular, our earlier results obtained for the hyperfine constants of the $3p\,^2P_{3/2}$ in atomic Na, found using a similar technique with expected similar systematics, agrees with the best of the large number of other measurements on that level \cite{NaHFQB}, but is more precise.  Beyond the statistical fluctuations in the measured polarization, the main source of potential uncertainty is in the calibration of the delay time.  However, it is quite straightforward to measure the length of a delay line of 30 m to within a cm or less. This corresponds to a time delay of 100 ns measured to within a fraction of a ns.

\begin{table}[h!]
\centering
\caption{Experimental results for the hyperfine coupling constants of the $8p\,^2P_{3/2}$ level of $^{133}$Cs.  Here, QBS refers to quantum beat spectroscopy, ODR indicates optical double resonance.  RMBT stands for relativistic many body theory.  Note that the A and B coefficients for this work have assigned 2$\sigma$ error bars.}
\begin{tabular}{ccccc}
\hline \hline $A$  &  $B$  & Technique & Sources \\ \hline
$7.42(6)$ & +0.14(29) & QBS & This work\\ $7.58(1)$  &
-0.14(5)  & ODR & \cite{faist64}\\  $7.626(5)$ &-0.049(42) &
ODR & \cite{bucka63}\\ $7.644(25)$&           & ODR &
\cite{abele75a,abele75b}\\ $7.27$ & & RMBT & \cite{Safronova1} \\
$7.55(5)$  & +0.63(35) & Model & \cite{barbey62}\\

\hline \hline
\end{tabular}
\end{table}

\section{Conclusions}
We have introduced a new technique to measure the hyperfine energy level separations in the 8p $^{2}P_{3/2}$ level of atomic Cs.  The approach is based on the influence of oscillating atomic multipoles in the excited states on the probe light polarization dependence of the stimulated emission rate to a lower energy level. We anticipate that this technique is applicable to measurements of small energy separations in many atomic systems.  It further should be extremely useful for measuring structural properties of diatomic molecules, particularly rotational energy separations and finer structure within rotational levels   (hyperfine structure and lambda doubling for instance).  The experimental approach, combination of pulsed optical pumping and time-delayed stimulated emission probing, has been used to measure the magnetic dipole (A) and electric quadrupole (B) hyperfine coupling constants in the atomic Cs 8p $^{2}P_{3/2}$ level.  Our results for A and B seem free of major systematic errors within the quoted uncertainties of the measurements.

\section{Acknowledgements} Financial support from the National Science Foundation (Grant No. NSF-PHY-1309571) is gratefully acknowledged.  One of us (MDH) also acknowledges financial support of the National Science Foundation (Grant No. NSF-PHY-1068159). The authors from Miami University also acknowledge Greg Reese, from Research Computing Support, for providing the Mathlab programming code to the data fitting.

\end{document}